\documentclass{svjour2}                    
\smartqed  
\usepackage{graphicx}
\usepackage{amssymb}

\def\smiley{\hbox{\large$\bigcirc$\hspace{-.80em}%
\raise.2ex\hbox{$\cdot\cdot$}\kern-.61em    
\lower.2ex\hbox{\scriptsize$\smile$}}\ }

\def\frowney{\hbox{\large$\bigcirc$\hspace{-.80em}%
\raise.2ex\hbox{$\cdot\cdot$}\kern-.635em
\lower.2ex\hbox{\scriptsize$\frown$}}\ }

\newcommand{\up}{\uparrow}
\newcommand{\dow}{\downarrow}
\newcommand{\caln}{{\cal{N}}}
\begin{document}

\title{On the Role of Entanglement in Schr\"{o}dinger's Cat Paradox
}


\author{Stefan Rinner        \and
        Ernst Werner 
}


\institute{S. Rinner \at
               Institut f\"{u}r Theoretische Physik, Universit\"{a}t Regensburg,
Germany\\
\email{stefan.rinner@physik.uni-regensburg.de}           
           \and
           E. Werner \at
Institut f\"{u}r Theoretische Physik, Universit\"{a}t Regensburg,
Germany \\Tel.: +49-941-9432001\\
              Fax: +49-941-9431734\\
              \email{ernst.werner@physik.uni-regensburg.de} }

\date{Received: date / Accepted: date}

\maketitle

\begin{abstract}
In this paper we re-investigate the core of Schr\"{o}dinger's 'cat
paradox'. We argue that one has to distinguish clearly between
superpositions of macroscopic cat states $|\smiley \rangle +
|\frowney \rangle$ and superpositions of entangled states
$|\smiley, \up \rangle + |\frowney, \dow \rangle$ which comprise
both the state of the cat ($\smiley$=alive, $\frowney$=dead) and
the radioactive substance ($\up$=not decayed, $\dow$=decayed). It
is shown, that in the case of the cat experiment recurrence to
decoherence or other mechanisms is not necessary in order to
explain the absence of macroscopic superpositions. Additionally,
we present modified versions of two quantum optical experiments as
{\sl experimenta crucis}. Applied rigorously, quantum mechanical
formalism reduces the problem to a mere pseudo-paradox.
\end{abstract}

\section{Introduction}
\label{intro}
Recently, there has been a number of reports on cooling of micromirrors \cite{Zeilinger}, \cite{Heidman} and micromechanical resonator \cite{Bouwemeester} down to such low temperatures that quantum effects such as superposition and entanglement at a macroscopic scale come into reach. Also, photoassociative formation of macroscopic atom-molecule superposition in Bose-Einstein-condensates has been considered lately theoretically \cite{Mackie}. Almost all works dealing with macroscopic superposition of one kind or another refer to the cat paradox claiming the cat itself being in a superposition state. Yet, as mentioned by Leggett \cite{Leggett} "the conceptual status of the theory is still a topic of lively controversy" and we would like to contribute to this controversy an alternative point of view which quite naturally explains the suppression of interference effects in macroscopic objects already at the level of isolated systems.\\
For the sake of completeness we briefly give the basic ingredients
of the {\it Gedankenexperiment}. The proposal involves a cat
(macroscopic), a vial of cyanide and a radioactive atom
(microscopic) initially prepared in a metastable state. All three
components are placed inside a closed box. The radioactive atom
has a probability of $1/2$ for decaying within one hour. If it
decays the cyanide shall be unharnessed and will kill the cat via
some mechanism. In Schr\"{o}dinger's own words
\cite{Schroedinger}:\begin{quotation}If one has left this entire
system to itself for an hour, one would say that the cat still
lives {\it if} meanwhile no atom has decayed. The first atomic
decay would have poisoned it. The $\psi$-function of the entire
system would express this by having in it the living and the dead
cat (pardon the expression) mixed or smeared out in equal parts.
\end{quotation}
The first sentence of this quotation emphasizes the entangled
 character of the system's state by stressing the word "if".\\
 In the third sentence
 Schr\"{o}dinger refers to the $\psi$-function of the {\bf entire}
 system. Forasmuch then as Schr\"{o}dinger neither claims the cat
 to be in a superposition state nor even uses the term 'paradox'
 throughout the article the succeeding interpretations of his
 {\it Gedankenexperiment} can only be thought of having misconstrued Schr\"{o}dinger's intention.
In fact, a paradox could only arise when from claiming the nucleus
to be in a superposition state one concludes that,
due to the entanglement between states of the atom and states of the cat, 
the cat is in a superposition of its two possible states, too,
{\it a minore ad maius}, so to speak. Exemplary for this attitude
we quote \cite{Burnett}:\begin{quotation}Quantum mechanics tells
us that at any time the nucleus involved is in a superposition of
the decayed and original state. Because the fate of the cat is
perfectly correlated with the state of the nucleus undergoing
decay, we are forced to conlude that the cat must also be in a
superposition state, this time of being alive and
dead.\end{quotation} This assessment is wide-spread in the
literature (see {\it e.g.}  \cite{Omnes}, \cite{Auletta} and
citations therein). In this paper we investigate a different
proposition that has the advantage of yielding non-paradoxical
predictions. Contrary to the statement cited above, it asserts
that at any time neither the nucleus (if the initial state is
eq.(\ref{Anfang})) nor the cat are in a superposition.\\ The fact
of the matter is that already three years before Schr\"{o}dinger's
article von Neumann treating the properties of composite systems
made the point clear \cite{Neumann}:\begin{quotation} Auf Grund
der obigen Resultate heben wir noch hervor: Ist I im Zustande
$\phi(q)$ und II im Zustande $\xi(r)$, so ist I+II im Zustande
$\Phi(q,r)=\phi(q)\xi(r)$. Ist dagegen I+II in einem Zustande
$\Phi(q,r)$, der kein Produkt $\phi(q)\xi(r)$ ist, so sind I und
II Gemische, aber $\Phi$ stiftet eine ein-eindeutige Zuordnung
zwischen den m\"{o}glichen Werten gewisser Gr\"{o}\ss{}en in I und
in II.
\end{quotation}
In English and contemporary diction, the main result of his
analysis of composite systems is the following: if a composite
system is in an entangled state, each of its subsystems is in a
mixed state. Thus, for the case in question here, the subsystem
'cat' is described by a mixed state, as well, and consequently is
not in a superposition state.

\section{Superposition and Entanglement}
\label{sec:1}

Since there is no correlation between cat and radioactive material
in the very beginning of the {\it Gedankenexperiment} the state
vector of the combined system may be written as a tensor product
in the following way
\begin{equation}
\label{Anfang} |\Psi(0) \rangle = |\up \rangle \otimes |\smiley
\rangle.
\end{equation}
In the course of time both subsystems become entangled and the
system's state can be written
\begin{equation}
\label{unter} | \Psi(t) \rangle = \Big(e^{-\frac{1}{2} \lambda
t}|\up, \smiley \rangle+\sqrt{1-e^{-\lambda t}}|\dow, \frowney
\rangle\Big).
\end{equation}
Two peculiarities of the given setup should be noted. First, the
Hilbert-space for the combined system is spanned by the four basis
states $\left\{|\up, \smiley \rangle, |\dow, \frowney  \rangle,
|\dow, \smiley  \rangle,  |\up, \frowney  \rangle \right\}$. Due
to the initial condition of eq.(\ref{Anfang}) only the subspace
spanned by the vectors given in eq.(\ref{unter}) is accessible.
Secondly, for $t>0$ the superposition of eq.(\ref{unter}) will
decay in time leading to a final state
$|\Psi(t \rightarrow \infty)\rangle=|\dow, \frowney \rangle$ even without the impact of an external environment.\\
The assumed half-life of one hour gives for the decay constant
$\lambda=\frac{\ln(2)}{3600 \:s}$ and for the corresponding state
after one hour
\begin{equation}
|\Psi'\rangle:=| \Psi(1 \:h) \rangle =
\frac{1}{\sqrt{2}}\Big(|\up, \smiley \rangle+|\dow, \frowney
\rangle\Big).
\end{equation}
Note that this statement is about the whole system being in a superposition state, but not concurrently a statement about the subsystems. In order to gain information about the state of subsystem {\bf A} of a combined system {\bf AB} the rules of quantum mechanics tell us that one should consider the system's density matrix rather than the state vector description and take the partial trace over the degrees of freedom of subsystem {\bf B}. \\
Thus, considering the density matrix of the evolved state
\begin{eqnarray}
\nonumber \hat{\rho'}_{sys.}&=&\frac{1}{2}\Big(|\up, \smiley
\rangle \langle  \smiley, \up|+|\dow, \frowney \rangle \langle
\frowney, \dow|+ \\ \nonumber &+&|\up, \smiley \rangle \langle
\frowney, \dow|+|\dow, \frowney \rangle \langle \smiley,
\up|\Big).
\end{eqnarray}
the density matrix describing the cat alone results from taking
the partial trace
\begin{eqnarray}
\nonumber
\hat{\rho'}_{cat}&=&\langle \up| \hat{\rho'}_{sys.} |\up \rangle + \langle \dow| \hat{\rho'}_{sys.} |\dow \rangle \\
&=& \frac{1}{2}\Big(|\smiley \rangle \langle \smiley|+|\frowney
\rangle \langle \frowney|\Big).
\end{eqnarray}
This means that within the framework of quantum mechanics there actually is no paradox, since the above reduced density matrix for the subsystem 'cat' is a statistical mixture of states {\sl dead} or {\sl alive} with equal probability $1/2$. The situation is the same as in classical statistics when one describes the unknown outcome ({\sl head} or {\sl tail}) of tossing a coin. No superposition state of the cat is present which would give rise to non-diagonal entries in the cat's density matrix $\hat{\rho'}_{cat}$.\\
At first sight, introducing the partial trace in such a way and declaring it a rule of quantum mechanics might seem just a clever trick in order to circumvent the interpretational difficulties posed by the paradox. Indeed, why should one choose to define the reduced state of a subsystem just in that way?\\
In order to justify this, consider a composite system {\bf AB}
whose state space is described by a tensor product of Hilbert
spaces ${\cal{H}}_{AB}={\cal{H}}_{A} \otimes {\cal{H}}_{B}$ with
${\cal{H}}_{A} \cap {\cal{H}}_{B}=\emptyset$. Then, if
${\cal{O}}_{A}$ is some observable of subsystem {\bf A} acting on
${\cal{H}}_{A} $ the corresponding observable acting on
${\cal{H}}_{AB}$ is consistently defined by
${\cal{O}}={\cal{O}}_{A} \otimes \hat{{\bf{1}}}_{B}$, where
$\hat{{\bf{1}}}_{B}$ is the identity-operator on ${\cal{H}}_{B}$.
When subsytem {\bf A} is prepared in a state described by
$\rho_{A}$ the expectation value of ${\cal{O}}_{A}$ should equal
the expectation value of ${\cal{O}}_{A} \otimes
\hat{{\bf{1}}}_{B}$ when we prepare the combined system in
$\rho=\rho_{A} \otimes \rho_{B}$. That is, consistency of
measurement statistics demands the following equality to hold:
\begin{equation}
Tr({\cal{O}}_{A} \rho_{A})=Tr([{\cal{O}}_{A} \otimes
\hat{{\bf{1}}}_{B}] \rho)
\end{equation}
It can be shown that this equation can only be satisfied if the
state of the subsystem $\rho_{A}$ is defined via the partial
trace:
\begin{eqnarray*}
\langle {\cal{O}} \rangle&=&Tr_{A,B}\Big(\rho {\cal{O}}\Big)=\sum_{a,b} \langle a,b| \rho {\cal{O}}|b,a \rangle= \\
&=&\sum_{a} \langle a| \sum_{b} \langle b| \rho
\hat{{\bf{1}}}_{B}|b \rangle {\cal{O}}_{A} |a \rangle=
Tr_{A}\Big(\rho_{A}{\cal{O}}_{A}\Big)=\\&=&\langle {\cal{O}}_{A}
\rangle
\end{eqnarray*}
This shows that in fact there is no freedom of choice in the way one defines the state of a subsystem.\\
Note, that in this deduction there was no assumption about the
size of the quantum subsystems. That is, one does not need to
emphasize the macroscopic size of the cat and interpret the cat
itself as some measurement apparatus or even to call for some sort
of consciousness of the cat. In particular, the same still holds
if the two subsystems are two-level systems like one atom with
states $|e \rangle$, $|g \rangle$ and the radiation field inside a
cavity with number states $|0 \rangle$ and $|1 \rangle$. If the
system is in a superposition state
\begin{equation}
\frac{1}{\sqrt{2}} \Big(|e,0 \rangle + |g,1 \rangle \Big)
\end{equation}
neither the atom nor the cavity field alone are in a
superposition. In the same line of reasoning and going from Fock
states further to coherent field states (Glauber states) of
mesoscopic size, the state in eq.(1) of \cite{Haroche} of the form
\begin{equation}
| \Psi \rangle=\frac{1}{\sqrt{2}} \Big(|e, \alpha e^{i \phi}
\rangle + |g, \alpha e^{-i \phi} \rangle \Big)
\end{equation}
actually does not describe a superposition of the coherent field states $|\alpha e^{i \phi} \rangle$ and $|\alpha e^{-i \phi} \rangle$. \\
Here, some words about change of basis seem to be in order. It is
clear that a mere rotation of axes will not change the situation,
{\it e.g.} consider the case of symmetric (S) and antisymmetric
(A) linear combinations of the "old" basis states defined in the
usual way:
\begin{eqnarray*}
|S \rangle &=& \frac{1}{\sqrt{2}}\Big(| \smiley \rangle + | \frowney \rangle\Big), \: \:|+ \rangle = \frac{1}{\sqrt{2}}\Big(| \up \rangle + | \dow \rangle\Big) \\
|A \rangle &=& \frac{1}{\sqrt{2}}\Big(| \smiley \rangle - |
\frowney \rangle\Big), \: \: |- \rangle = \frac{1}{\sqrt{2}}\Big(|
\up \rangle - | \dow \rangle\Big).
\end{eqnarray*}
Then
\begin{equation}
\frac{1}{\sqrt{2}}\Big(| \smiley, \up \rangle + | \frowney, \dow
\rangle\Big)=\frac{1}{\sqrt{2}}\Big(|S,+ \rangle + |A,-
\rangle\Big).
\end{equation}
Again, after tracing over the states $|+ \rangle$ and $|- \rangle$
the reduced density matrix is given by
\begin{equation} \nonumber
\hat{\rho'}_{cat}=\frac{1}{2}\Big(|S \rangle \langle S|+|A \rangle
\langle A|\Big)=\frac{1}{2}\Big(|\smiley \rangle \langle \smiley
|+ |\frowney \rangle \langle \frowney |\Big)
\end{equation}
where the last equality is gotten by transforming $|S \rangle$ and $|A \rangle$ back to the "old" basis states. So, in both bases the reduced density matrix is diagonal which does not pose any kind of interpretational problem.\\
Rather one could object that in quantum mechanics a measurable
state should be an eigenstate of some observable (operator) and
for the (anti-)symmetric combination-states a corresponding
observable could be difficult to define. At least, it is not
obvious what this would look like. As, by the way, is already the
case for the "alive" and "dead" states introduced by
Schr\"{o}dinger. In the latter case, it is easy to find some
alternative system to replace the cat, {\it e.g.} some mass
suspended on a thread that is cut if the atom decays. Thus, the
macroscopically distinct states $| \top>$(=mass hangs on the
thread) and $|\bot \rangle$ (=mass fallen to the floor) correspond
to the liveliness of the cat. If the thread is attached at some
place outside the box and screened from view in one way or the
other, the weight of the box could be an appropriate observable
that allows one to discriminate both states from each other.

\section{Experimental Test}
\label{sec:2} Whether the reduced density matrix of a subsystem is
enough in order to describe the state of the subsystem completely
and correctly, should not be a question of philosophical taste but
should be decided at first instance by experiments.
\subsection{Paris experiment}
\label{sec:3} The first of such tests consists in the modification
of an experiment Brune {\em et al.} \cite{Haroche} have carried
out in quantum optics. It constitutes the experimental adoption of
an earlier theoretical proposal by Schaufler {\em et
al.}\cite{Schleich}. The setup substantially is made up by a
high-Q microwave resonator $C$ containing a coherent field
$|\alpha \rangle$ and Rydberg atoms with excited $|e \rangle$ and
ground state $|g \rangle$ that are used both to manipulate and
probe the field. Thereto, before entering C the atom is prepared
in an superposition of $|e \rangle$ and $|g \rangle$ in low-Q
cavity $R_1$ by a resonant $\pi/2$ pulse. This superposition state
enters $C$ and being detuned from resonance interacts dispersively
with the cavity field in $C$. This interaction produces an
atom-level depending phase shift of the cavity field and leads to
the following entangled atom-field state:
\begin{equation}
\label{inkoh} | \Psi \rangle_{R_1C}=\frac{1}{\sqrt{2}}\Big(|e,
\alpha \rangle + |g, -\alpha  \rangle\Big)
\end{equation}
where the subscript on the left hand side indicates that the atom has already passed $R_1$ and $C$. Now, following the widespread opinion one would say that the field is already in a superposition.\\
Contrary to accepted opinion, we hold a different view that is
based on the importance we ascribe to the reduced density matrix.
After leaving $C$ the atom undergoes another $\pi/2$ pulse in a
second resonator $R_2$ leading to the system state:
\begin{equation}
\label{koh} | \Psi \rangle_{R_1CR_2}=\caln \left[\Big(|-\alpha
\rangle -|\alpha \rangle \Big)|e \rangle + \Big(|-\alpha \rangle
+|\alpha \rangle \Big)|g \rangle\Big) \right]
\end{equation}
with some normalisation constant $\caln$. Behind $R_2$ the atom is detected state-selectively. This projects the field state onto $|\alpha  \rangle +e^{i \psi}|\alpha \rangle$  with $\psi=0$ or $\psi=\pi$, according to whether the state of the atom was  $|g \rangle$ or $|e \rangle$ respectively.\\
In the original version of the experiment this state is
susequently probed by a second atom that is sent into the setup
after a variable time interval $\tau$ in order to monitor
decoherence. The signature of progressive loss of coherence is the
decay of the two-atom correlation signal as a function of the
preparation-probing interval $\tau$.\\ If eq.(\ref{inkoh}) already
was a Schr\"{o}dinger cat state, as argued usually, the second
Ramsey-zone in the setup of Brune {\em et al.} would have been
needless. Indeed, the quantum interference signal is explained
through erasing {\it welcher-Weg-}information in
\cite{Davidovich}. Therefore, repeating measurements on this
apparatus and leaving out the interaction in $R_2$ in the
preparation process as well as the state-selective detection of
the preparing atom could decide whether one could measure the
two-atom correlation signal at all in this modified version.
\subsection{Garching experiment}
\label{sec:4}
In this paragraph we propose a modification of another experimental setup in order to show that the absence of quantum interference in subsystem-states when the entangled system is in a superposition is not a peculiarity of one of the subsystems being (quasi) macroscopic as in eq.(\ref{inkoh}). A preparation scheme for superposition states of highly non-classical photon number states of a radiation field inside a high-Q cavity was proposed in \cite{Rinner1} for one cavity and in \cite{Rinner2} for two coupled micromasers. Hereto, the coherent exchange of energy between Rydberg atoms sent through the cavity with the radiation field is used. Intriguingly, following only the Rabi oscillations in the Jaynes-Cummings model of quantum optics quantum interference effects will be observable only if both the atom and the field are in coherent superpositions at the beginning of the interaction.\\
Suppose we start with the atom in the excited state and the cavity
mode in the vacuum:
\begin{equation}
|\psi(0) \rangle=|e \rangle |0\rangle.
\end{equation}
The time structure of the Rabi oscillation leads to
\begin{equation}
|\psi(t) \rangle=\cos(g t)|e,0 \rangle+i \sin(g t) |g,1 \rangle
\end{equation}
where $g$ denotes the vacuum Rabi frequency. At some point of the
Rabi oscillation two lasers are applied that induce transitions of
the atom from both state $|e \rangle $ and state $|g \rangle$ to
one and the same lower lying state $|a \rangle$. Note that $|g
\rangle$ is some highly excited Rydberg state and only the ground
state of the maser transition, not the "real" ground state of the
atom. Hereby, the {\it welcher-Weg-}information was erased giving
rise to observable quantum interference effects since the state of
the system after this procedure is given by
\begin{equation}
|\psi(t') \rangle = \Big( \cos(gt')|0 \rangle + i \sin(gt')|1
\rangle \Big)|a \rangle.
\end{equation}
Detection of the atom in $|a \rangle$ leaves the field in a coherent superposition.\\
This suggests that the general recipe for the generation of
(particularly macroscopic) superposition states is the following:
at first, create entanglement with another (microscopic) system.
This leads to superpositions of entangled states. Yet, in order to
transfer the coherence to one of the subsystems alone one has to
deliberately disentangle the two systems by erasing the {\it
welcher-Weg-}information in one subsystem and thus enabling
quantum interference effects in the other one.

\section{Measurement Problem}
\label{sec:5} Since any physical property finally has to be
measured in order to gain information about its value the act of
measurement plays a decisive role both in the formulation and
interpretation of theories in physics. The proposed interpretation
in terms of density matrices and partial trace operations for
subsystems of composed systems also obviates the so called
measurement problem (at least the part of it dealing with the
problem of definite outcomes). If the measurement device $\cal{M}$
allows for two readings $|\nearrow \rangle, \: |\nwarrow \rangle$
correlated with the cat's state of liveliness then the extended
density matrix reads
\begin{eqnarray*}
\hat{\rho}&=&\frac{1}{2}\Big(|\nwarrow, \up, \smiley \rangle
\langle  \smiley, \up, \nwarrow|+|\nearrow, \dow, \frowney \rangle
\langle \frowney, \dow, \nearrow|+
\\&+&|\nwarrow, \up, \smiley \rangle \langle \frowney, \dow,
\nearrow|+|\nearrow, \dow, \frowney \rangle \langle \smiley, \up,
\nwarrow|\Big).
\end{eqnarray*}
which gives for the measurement device's reduced density matrix
\begin{equation}
\hat{\rho}_{{\cal{M}}}=\frac{1}{2}\Big(|\nwarrow \rangle \langle
\nwarrow|+|\nearrow \rangle \langle \nearrow|\Big).
\end{equation}
The same is still true if the measurement apparatus allows for
more than two pointer states as in the original formulation of the
problem by von Neumann \cite{Neumann}. There, one considers a
(microscopic) system ${\cal{S}}$ with Hilbert space ${\cal{H_S}}$
and basis vectors $|s_n \rangle$ together with a (macroscopic)
measurement apparatus ${\cal{A}}$ with Hilbert space ${\cal{H_A}}$
and basis vectors $|a_n \rangle$ that are supposed to correspond
to macroscopically distinguishable pointer states. Further, it is
assumed that a pointer reading of $|a_n \rangle$ corresponds to
the state $|s_n \rangle$ of system ${\cal{S}}$. If $|a_0 \rangle$
denotes the ready-position of the apparatus the following
evolution will take place:
\begin{equation}
\left( \sum_{n} c_n(0) |s_n \rangle \right)|a_0 \rangle
\longrightarrow \sum_{n} c_n(t) |s_n \rangle |a_n \rangle.
\end{equation}
The reduced density matrix of the apparatus has only diagonal
entries:
\begin{equation}
\left( \hat{\rho}_{\cal{A}}\right)_{nn}(t)=|c_n(t)|^2
\end{equation}
Consequently, the outcomes of measurements are statistically
distributed, yet definite ones.

\section{Conclusion}
\label{sec:6}
First, in the preceeding it was shown that neither decoherence, {\it i.e.} entanglement with some environment, nor other ideas like superpositions of space-time geometries \cite{Penrose} need to be invoked in order to arrive at a classical picture of the Schr\"{o}dinger cat scenario. This shall not derogate the clarification on the role of the environment accomplished by the decoherence program. In the case in question here, resorting to decoherence in order to arrive at classicality of the cat is not necessary and still features interpretational problems in explaining {\it e.g.} how small the off-diagonal elements of the density matrix must be in order to call the density matrix a statistical mixture since they vanish only in the limit $t \longrightarrow \infty $. Hence, the transition from the (alleged) macroscopic superposition state to the familiar statistical mixture would still necessitate the existence of some observer and would depend on his ability to resolve the "distance" of the individual components of the superposition state.\\
Quite contrary, we argue that for the generation of (macroscopic) superposition states of the Schr\"{o}dinger cat kind some initial entanglement with a microscopic system has to be removed from the composite system later on by performing a transformation on the microscopic system that erases the {\it welcher-Weg-}information. In fact, the signature of coherent superposition states is the interference pattern of some proper measureable quantity. This interference arises if the system starting out from its initial state A has two or more possibilities B, C, D ... to end up in one final state Z. Yet, in the case of Schr\"{o}dinger's cat there is no such final state to which two or more different paths would have been open. Where, then, should interference come from? \\
Secondly, it is more satisfying to have  a self-consistent interpretation which does not contradict everyday experience ({\it i.e.} no superposed cats), but still is able to fully reproduce measurements performed on intentionally prepared superposition states, exemplarily shown for \cite{Haroche}, \cite{Rinner1}. This is guaranteed by interpreting the reduced density matrix as a quantity that completely describes that state of a given subsystem. \\
To summarize, Schr\"{o}dinger's cat paradox in our opinion has its
roots in the state vector description of the composite system
which indeed shows a superposition of states (i.e. entangled
state). In order to come from this entangled state to a
superposition of the two cat states one has to ignore the
non-identity of the two nuclear states. This step is much less
innocent than it might look: it changes - by hand - an entangled
state into a coherent state. In physical reality such a
transmutation could only be achieved by erasing
"which-path-information" ( see the discussion in sec. 3.2) which
is not the case here.\\
We have shown that, at least in the two cases discussed above, a well-chosen transformation on one of the subsystems can lead to a disentangled state which indeed leaves the other subsystem in a superposition state.\\
Although the interpretation of the mathematical formalism
underlying a physical theory to a certain extent has a right in
its own, it holds the dangerous tendency to misconceive itself as
the 'philosophy of nature' in the sense that the elements of the
theory are taken to correspond to essential properties of reality.
Interpretations of quantum mechanics are particularly prone to
this ontological persuasion. Yet, the relation between the
formalism and the supposedly underlying reality it tries to
describe cannot be treated within the formalism itself. Since
physical theories are not part of the objects investigated by
quantum mechanics, quantum mechanics itself is not an object the
theory makes statements about. The connection between theory and
reality has to be established axiomatically in the formulation
of the theory and the theory has to be checked for consistency henceforth. \\




\end{document}